\documentclass{moriond}

\def\be{\begin{equation}}
\def\ee{\end{equation}}
\def\bea{\begin{eqnarray}}
\def\eea{\end{eqnarray}}

\newcommand{\Photo}{\includegraphics[height=35mm]{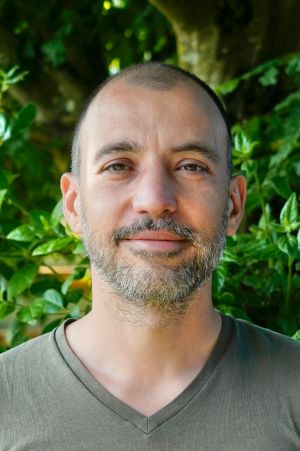}}

\begin{document}
\vspace*{4cm}
\title{Synergy between the Cherenkov Telescope Array Observatory\\ and the Vera C. Rubin Observatory}

\author{J. Biteau$^{1,2}$, J. Hamo$^{1}$, A. Mikhno$^{1,3}$, J. Peloton$^{1}$, J.-P. Lenain$^{3}$}

\address{$^{1}$ Université Paris-Saclay, CNRS/IN2P3, IJCLab, Orsay, France\\
$^{2}$ Institut universitaire de France (IUF), France\\
$^{3}$ Sorbonne Université, CNRS/IN2P3, Laboratoire de Physique Nucléaire et de Hautes Energies, LPNHE,
4 place Jussieu, 75005 Paris, France\\ \
}

\maketitle\abstracts{
The Cherenkov Telescope Array Observatory (CTAO) and the Vera C. Rubin Observatory are set to transform our understanding of the universe over the next decade.  These two observatories have multiple areas of complementarity in their scientific applications, ranging from constraints on cosmological parameters to studies of asteroid occultations. The most opportune area of synergy probably lies in the field of time-domain astronomy. Due to their sensitivity and saturation limits, it will be difficult for the two observatories to conduct joint studies of variable and transient sources in the Milky Way. However, they could offer a fresh and rich perspective on non-thermal extragalactic sources, in particular gamma-ray bursts, active galactic nuclei and jetted tidal disruption events. Among these sources lie the best candidates for multi-messenger research into the origin of TeV-PeV neutrinos and multi-EeV cosmic rays. Thus, combined with multi-wavelength observations by X-ray satellites and wide-field gamma-ray instruments, the synergy between Rubin and the CTAO could provide answers to some of the most important questions in astroparticle physics. This scientific potential comes with a challenge: selecting a few alerts from the ten million issued by Rubin each night to repoint the CTAO telescopes. We use the variability of blazars over timescales ranging from a few days to several years as a case study to demonstrate how to address this challenge using the Fink broker of Rubin.}

\section{Introduction}

Over the past two to three decades, deep observations of the sky have radically changed our understanding of the universe and its baryonic matter content. Consider, for instance, galaxy counts — the field of research that involves identifying galaxies in large surveys and reconstructing their flux distribution to understand star formation in galaxies and accretion processes around supermassive black holes across cosmic epochs. Pioneering studies in the optical and near-infrared bands at the beginning of this century, particularly those conducted with the Hubble Space Telescope, enabled the radiative balance of the post-recombination universe to be estimated this way with an uncertainty of around 40\%.\cite{2000MNRAS.312L...9M} Legacy data from this telescope, along with new deep observations provided by the James Webb Space Telescope, now make it possible to determine this radiative balance in optical bands with an accuracy of about 2\%,\cite{2026MNRAS.547ag044T} a finding supported by direct measurements from the outer edges of the Solar System and indirect measurements based on absorption techniques.\cite{2024ApJ...975L..18G} This revolution in measurement precision opens up new dimensions in the observable parameter space. Behind the integral and relatively static nature of galaxy-count observations lies the incessant variability of the most extreme astrophysical sources. These sources can appear and then disappear, or experience flux increases or decreases by several orders of magnitude over timescales ranging from seconds to decades. The observation and study of these sources falls within the field of time-domain astronomy, an area that the Vera C. Rubin Observatory promises to transform at optical wavelengths.

Over the same period, new methods for observing extreme sources in the universe emerged in the field of astroparticle physics. This area gained prominence with the completion of the Pierre Auger Observatory for cosmic rays at ultra-high energies (${>}\,10^{18}\,\mathrm{eV} \equiv 1\,\mathrm{EeV}$), the IceCube and KM3NeT observatories for neutrinos at  very-high energies (${>}\,10^{12}\,\mathrm{eV} \equiv 1\,\mathrm{TeV}$), as well as Cherenkov telescopes and the \textit{Fermi} satellite for gamma rays at very-high and high energies (${>}\,10^{9}\,\mathrm{eV} \equiv 1\,\mathrm{GeV}$), which will soon be observed with the Cherenkov Telescope Array Observatory (CTAO). Not to mention, of course, the gravitational wave observatories that are often associated with this field. Although gamma-ray sources usually have multi-wavelength counterparts that enable us to identify the nature of the accelerator, the origin of the highest-energy neutrinos and cosmic rays is still a matter of debate. The high magnetic luminosity associated with acceleration in such environments \cite{2009JCAP...11..009L} suggests that the most energetic particles in the universe are produced by variable and transient sources.\cite{2024ApJ...972....4M}

Against the backdrop of the commissioning of Rubin and the CTAO, we were invited to explore the synergies between these new major optical and gamma-ray observatories at the 60th Rencontres de Moriond conference.

\section{Rubin and CTAO: performance and science cases}

\begin{figure}
\centerline{\includegraphics[width=0.675\linewidth]{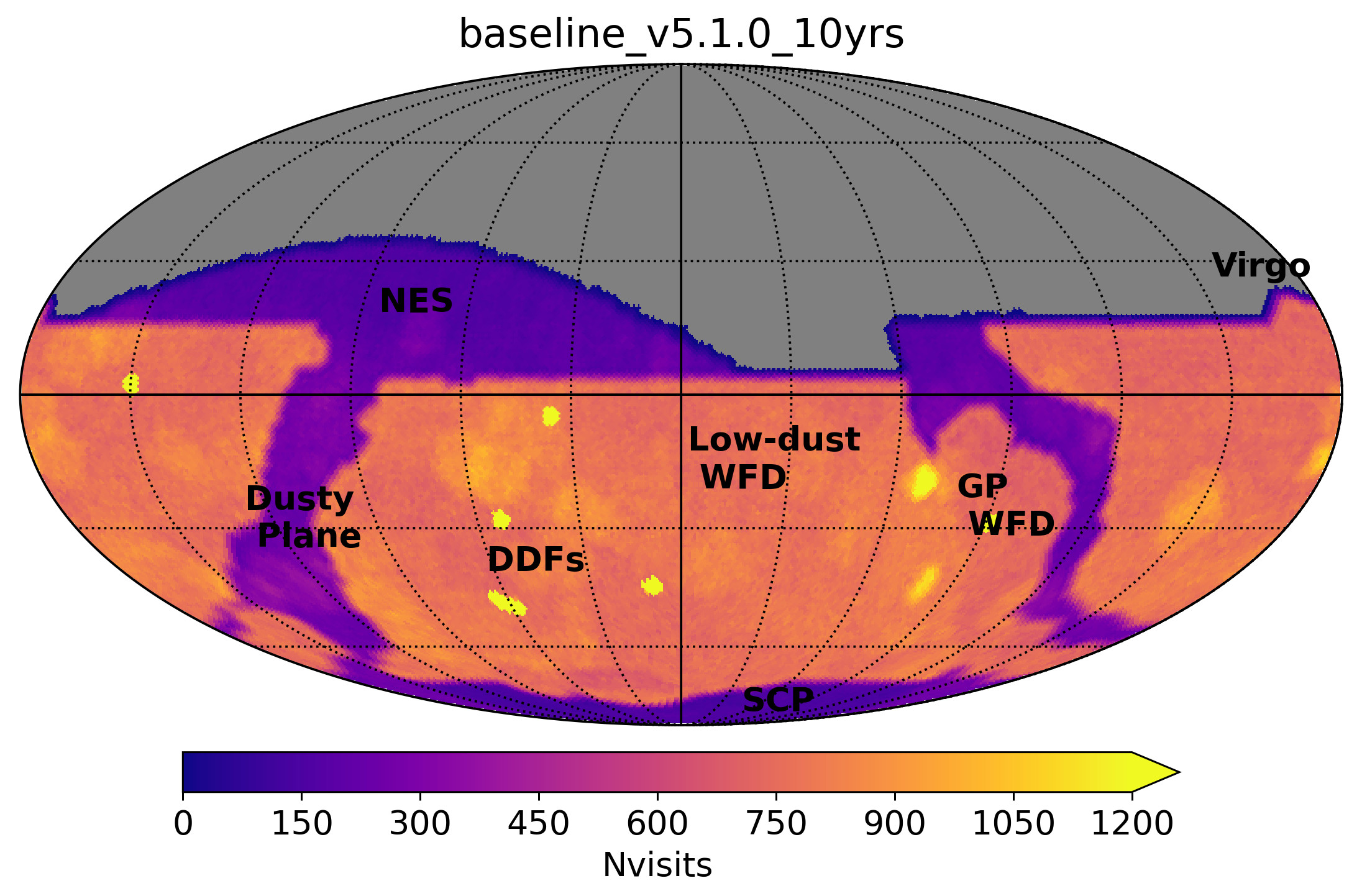}}
\caption[]{The expected footprint of Rubin's LSST (\texttt{v5.1.0}) in equatorial coordinates. The colour scale shows the number of visits per pixel in each band of Rubin after 10 years of observations. \textit{Adapted from Ref.~\cite{2026arXiv260121769P}}.}
\label{fig:LSSTfootprint}
\end{figure}

The Rubin Observatory, which began operations in 2025 at Cerro Pachón in Chile, is a six-band (ugrizy) optical observatory covering a wavelength range of approximately $0.3–1.1\,\mu$m. It features the largest CCD camera ever built, comprising approximately 3.2 billion pixels covering a field of view of about 3\,deg (9.6\,deg$^2$). This wide field of view will enable Rubin to conduct its Legacy Survey of Space and Time (LSST), by scanning the entire southern sky in each of its six bands up to the 27th magnitude after ten years of observations. Figure~\ref{fig:LSSTfootprint} illustrates the distribution of observations across the sky during the ten-year period. Each 30-second observation, with a cadence of approximately three days, will achieve a limiting magnitude of around 24, compared to a limiting magnitude of around 20 for the precursor survey conducted by the Zwicky Transient Facility (ZTF) in the Northern Hemisphere. Such a high sensitivity is necessarily accompanied by a saturation limit, of approximately magnitude 15.5 for Rubin,\cite{2009arXiv0912.0201L} versus magnitude 12.5 for ZTF.\cite{2019PASP..131a8002B} For reference, a microquasar in the Milky Way such as SS\,433 and a nearby jetted active galactic nucleus such as Mrk 501 typically have V-band magnitudes in the range 13–14 and are therefore bound to saturate Rubin’s CCDs most of the time. It is therefore among the fainter sources and in the low-surface-brightness universe that Rubin's full potential will be realised. The four pillars of Rubin's science lie in the study of small solar system objects (asteroids and comets), characterising dark energy properties (through SN\,Ia observations) and the distribution of dark matter (including microlensing and identifying new dwarf galaxies), studying the formation and evolution of the Milky Way (by observing more than ten billion stars) and surveying the variable and transient sky. In this regard, it should be noted that an increased cadence in a few deep drilling fields (DDFs, see Fig.~\ref{fig:LSSTfootprint}) could allow intranight variability to be investigated in a subsample of sources.\cite{2026arXiv260121769P}

Operating at wavelengths at least ten orders of magnitude shorter than those observed by Rubin, the CTAO is the new generation of ground-based gamma-ray observatories, covering an energy range from 20\,GeV to 300\,TeV. The Observatory is distributed across two sites: Paranal in Chile in the Southern Hemisphere and La Palma in the Canary Islands in the Northern Hemisphere. This allows the CTAO to access any variable source in the night sky. CTAO's alpha configuration comprises 13 telescopes at the northern site and 51 at the southern site. The telescopes will be deployed gradually, with 2027 marking the year when effective commissioning is expected to enable each array to surpass the sensitivity of current Cherenkov telescopes. The telescopes at each site, which vary in size, consist of large reflectors (4.3\,m to 23\,m in diameter) that focus the Cherenkov light flashes from gamma-ray atmospheric showers onto ultra-fast cameras. Each camera comprises approximately 2,000 photodetectors with GHz sampling. The optical design enables CTAO's cameras to observe a wide field of view: 4° for the large-sized telescopes and 8° for the small and medium ones. Extended regions of the sky, such as the Galactic plane or a portion of the extragalactic sky, can be scanned as a result. Its sensitivity, increased by nearly an order of magnitude compared to current Cherenkov telescopes, coupled with improved angular and energy resolution, will make the CTAO a formidable discovery machine. The main scientific objectives are to investigate the nature of dark matter — in the Galactic centre, galaxy clusters and dwarf galaxies — to understand the feedback role of cosmic rays — in the Milky Way, within star-forming galaxies or via jets from active galactic nuclei — and to explore extreme environments — by probing plasma near compact objects and constraining the propagation of gamma rays in cosmic voids. These propagation processes connect the CTAO to cosmology by opening a window onto parameters such as the Hubble constant, which gamma-ray cosmology now constrains in a manner competitive with gravitational-wave astronomy.\cite{2024ApJ...975L..18G} CTAO’s ultra-fast cameras could also open new horizons in optical astronomy through kHz-sampling of occultation events by asteroids or stellar intensity interferometry on kilometre-scale baselines.\cite{CTAConsortium:2017dvg}

\begin{figure}
\centerline{\includegraphics[width=0.675\columnwidth]{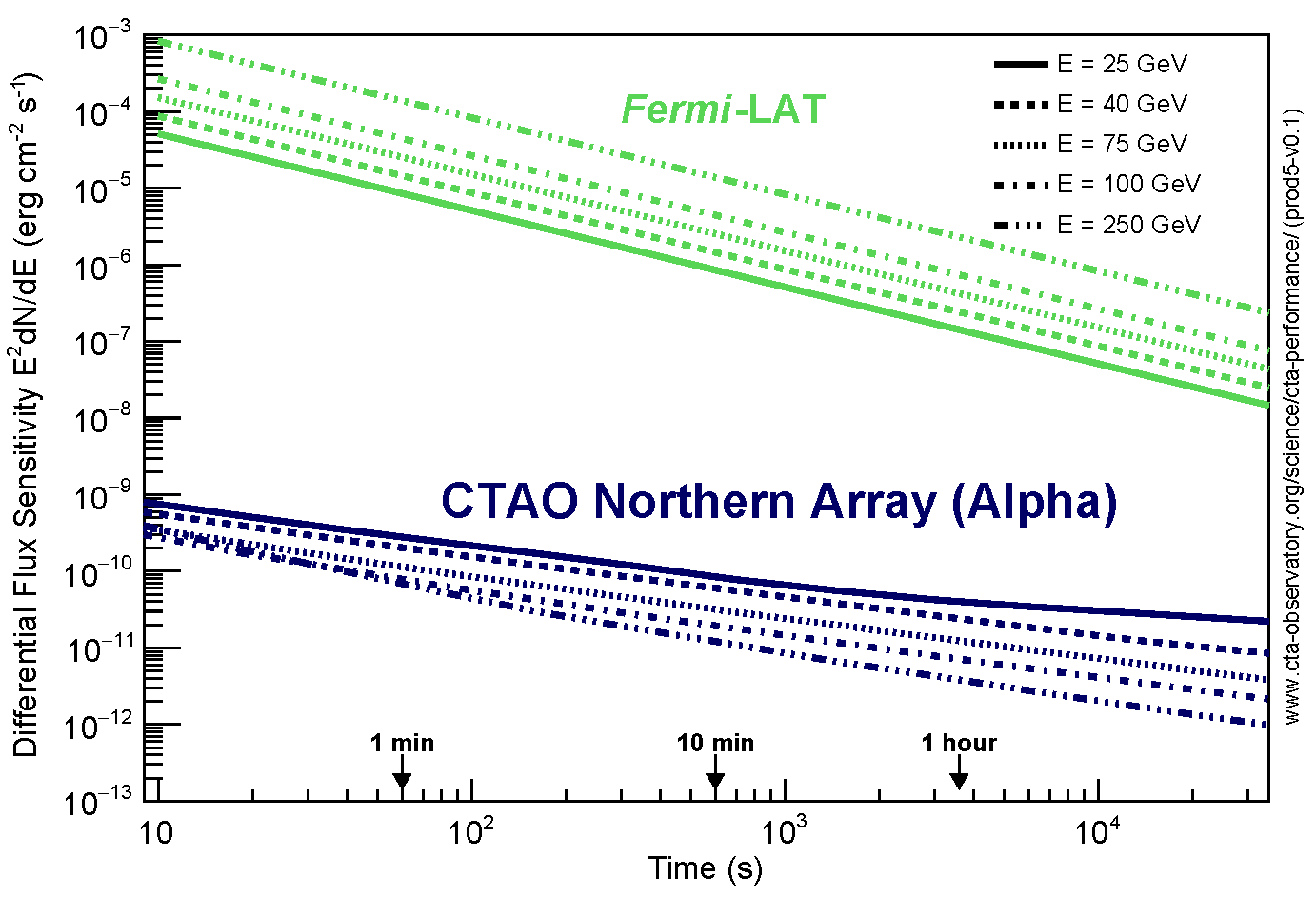}}
\caption[]{The sensitivity of the \textit{Fermi}-LAT satellite and of the CTAO-North array (alpha configuration \texttt{prod5-v0.1}) over timescales ranging from 10\,s to 10\,h. Sensitivity is expressed as the minimum energy flux detected at the $5\sigma$ confidence level in energy bins with a width of 0.2\,dex. The central bin energies range from 25\,GeV to 250\,GeV, as labelled in the figure. \textit{See Ref.~\cite{ctao_perf}}.}
\label{fig:CTAOtime}
\end{figure}

CTAO's coverage of baselines ranging from tens of metres to few kilometres, combined with large optical reflectors, provides it with an unprecedented effective collection area for gamma-ray energies above tens of GeV. Thus, whereas the collection area of an instrument such as the Large Area Telescope (LAT) onboard the \textit{Fermi} satellite is just under 1\,m$^2$, the collection area of each of the CTAO-North and CTAO-South arrays exceeds 10$^5$\,m$^2$ at around 100\,GeV. This technical characteristic will enable the CTAO to investigate non-thermal astrophysical processes on much shorter timescales than the \textit{Fermi}-LAT, as illustrated in Fig.~\ref{fig:CTAOtime}. It is in these time-domain capabilities, at timescales ranging from a few tens of seconds to a night, that the main synergy between the CTAO and the Rubin Observatory likely lies.

While a small fraction of Rubin's time (${\sim}\,3\%$) will be dedicated to reacting to external alerts, its survey of the transient sky will issue alerts every $30-60$ seconds to indicate points in the sky that differ at the $5\sigma$ level from a reference template. The reference template is constructed by averaging prior observations, typically of a few months' duration. Once this regime is reached, Rubin will emit five to ten million alerts per night, covering all types of events, including transient and variable sources, as well as bodies passing in front of the camera, such as asteroids. 

How will we select the few relevant alerts per night to point an observatory such as the CTAO from 2027 onwards? The processing, cross-matching with catalogues, classification, and redistribution of relevant alerts will be the responsibility of the seven brokers selected by Rubin.\cite{2025CoSka..55b..95V} In the following section, we explore the use of one of these brokers, Fink, which aims to address all of Rubin's scientific cases in a general manner by offering the community the ability to develop information processing modules, as well as filters for selecting the subset of alerts that are of interest to them. We demonstrate the relevance of this broker for a use case involving the dominant class of gamma-ray sources in the extragalactic sky: blazars. We use ZTF data as a precursor to Rubin for optical observations. For gamma-ray observations, we use \textit{Fermi}-LAT data as an unbiased survey of the sky that the CTAO will observe below 100\,GeV.

\section{Use case: variability of blazars}

\begin{figure}
\centerline{\includegraphics[width=0.99\linewidth]{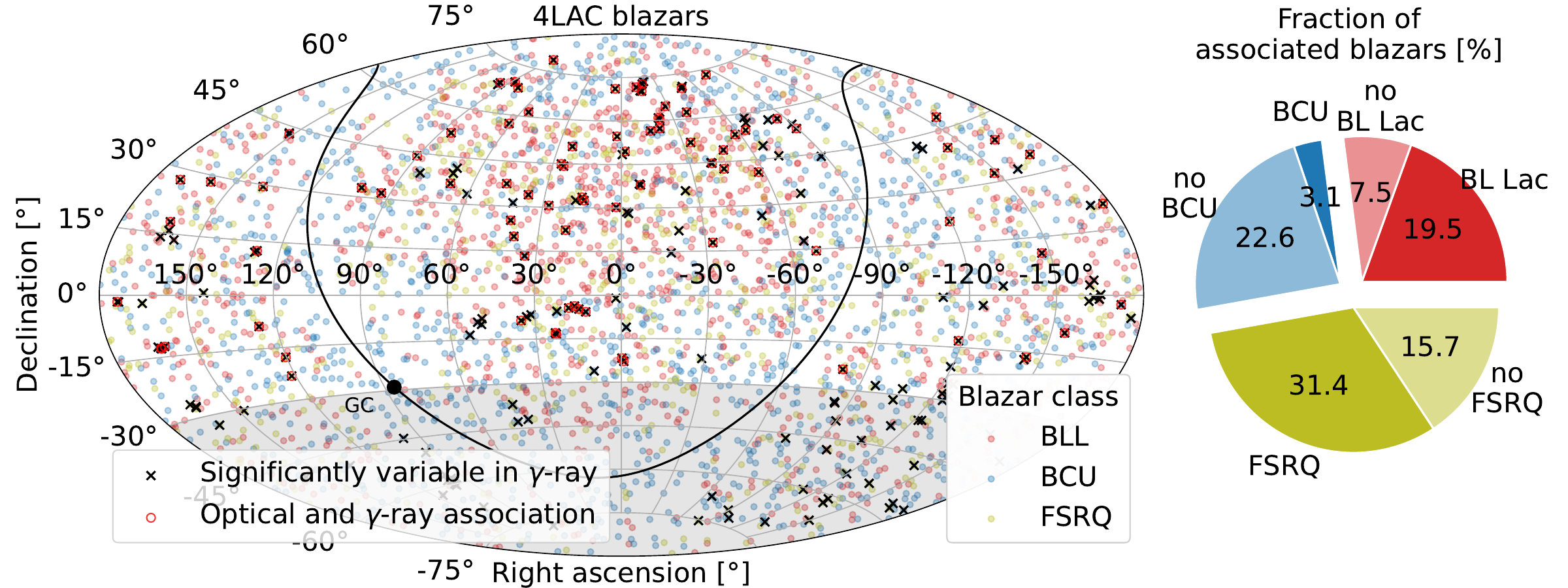}}
\caption[]{\textit{Left:} The skymap, in equatorial coordinates, of gamma-ray emitting blazars from the \textit{Fermi}-LAT 4LAC-DR2 catalogue,\cite{2020ApJ...892..105A} cross-matched with the ZTF optical survey (data release 23).\cite{2020PASP..132c8001D} Gamma-ray blazars with a fractional variability on weekly timescales larger than 30\% and significant at the $5\sigma$ level are marked with crosses. Associated optical lightcurves from ZTF with more than 150 measurements and a fractional variability larger than 10\% are shown with red circles. The colour of the markers subdivides the blazars into BL Lacs (BLL), flat-spectrum radio quasars (FSRQs) and blazars of an unknown class (BCU). The grey-shaded area shows the declinations that are inaccessible to ZTF. The black line and circle indicate the Galactic plane and centre (GC), respectively. \textit{Right}: The pie chart of selected variable gamma-ray blazars, with or without an optical counterpart as shown with darker and lighter colours, respectively. Colours display the blazar class. \textit{Adapted from Ref.~\cite{HamoInPrep}}.}
\label{fig:ZTFxLAT}
\end{figure}

Blazars are active galactic nuclei with a plasma jet oriented close to the line of sight. These objects are rare among the population of galaxies accessible via large optical surveys: only a few percent of red (elliptical) galaxies have an active nucleus in the local universe ($z<1$),\cite{2017A&A...601A..63W} only 10\% of them are classified as radio-loud, producing a double relativistic jet with a bulk Lorentz factor $\Gamma \approx 10$,\cite{2019ARA&A..57..467B} and only a few per thousand to a few percent of these jets are sufficiently aligned with the line of sight (cone of half-angle 1/$\Gamma$) to be observed as blazars. However, these jets are sites of tremendous particle acceleration, and their alignment boosts the observed energy and intensity of the particle radiation via the relativistic Doppler effect. These properties have enabled nearly 3,000 gamma-ray blazars to be detected over 10~years of \textit{Fermi}-LAT observations between 50\,MeV and 1\,TeV,\cite{2020ApJ...892..105A} as illustrated in Fig.~\ref{fig:ZTFxLAT} . 

The observed proportion of active galactic nuclei in red galaxies increases sharply with redshift, reaching almost 25\% at cosmic noon ($z \approx 2$).\cite{2017A&A...601A..63W} Those with jets can now be detected in GeV gamma rays up to $z \approx 5$ during their most powerful outbursts.\cite{2025ApJ...990..206G} The origin of these outbursts and of the variability of jetted active galactic nuclei remains an unresolved question to this day. Studying the variability of blazars across multiple timescales (from decades to minutes) and at multiple wavelengths could be key to understanding the link between accretion and ejection around supermassive black holes. Furthermore, the observation of bright blazar outbursts in TeV gamma rays provides a unique tool for probing the extragalactic background light mentioned in the introduction, magnetic fields in the intergalactic plasma, as well as physics beyond the Standard Model such as Lorentz invariance violation or coupling to axion-like particles.\cite{2022Galax..10...39B} So far, the observation of such outbursts with CTAO predecessors has relied mostly on the high-flux states of \textit{Fermi}-LAT,\cite{Lenain:2017okf} particularly for the most distant TeV blazars seen today up to $z \approx 1$.

Following seminal studies of blazars observed at optical wavelengths and in GeV gamma rays, we select in Ref.~\cite{HamoInPrep} a sample of 86 blazars with sufficient temporal coverage over the past seven years and sufficient variability in both bands to explore correlations. These 86 blazars are circled in red in Fig.~\ref{fig:ZTFxLAT}. Blazars show a non-thermal broadband emission spectrum, so the optical emission in the g and r bands of ZTF carries comparable information. To improve the effective cadence of optical observations, we combine the two ZTF optical bands in the form of a standardised time series, obtained by dividing each flux light curve by the median over same-hour observations. This standardisation process is most appropriate when non-thermal emission dominates that of the host galaxy. Otherwise, it could introduce distortions to the light curve. As we find only mild residual intra-night colour variations, less than 40 milli-mag RMS, we can confidently conclude that any bias introduced by our method does not affect the study of Ref.~\cite{HamoInPrep}. The standardisation process, applicable to blazars observed by ZTF and Rubin, is implemented in a dedicated Fink module. 

\begin{figure}
\centerline{\includegraphics[width=0.97\linewidth]{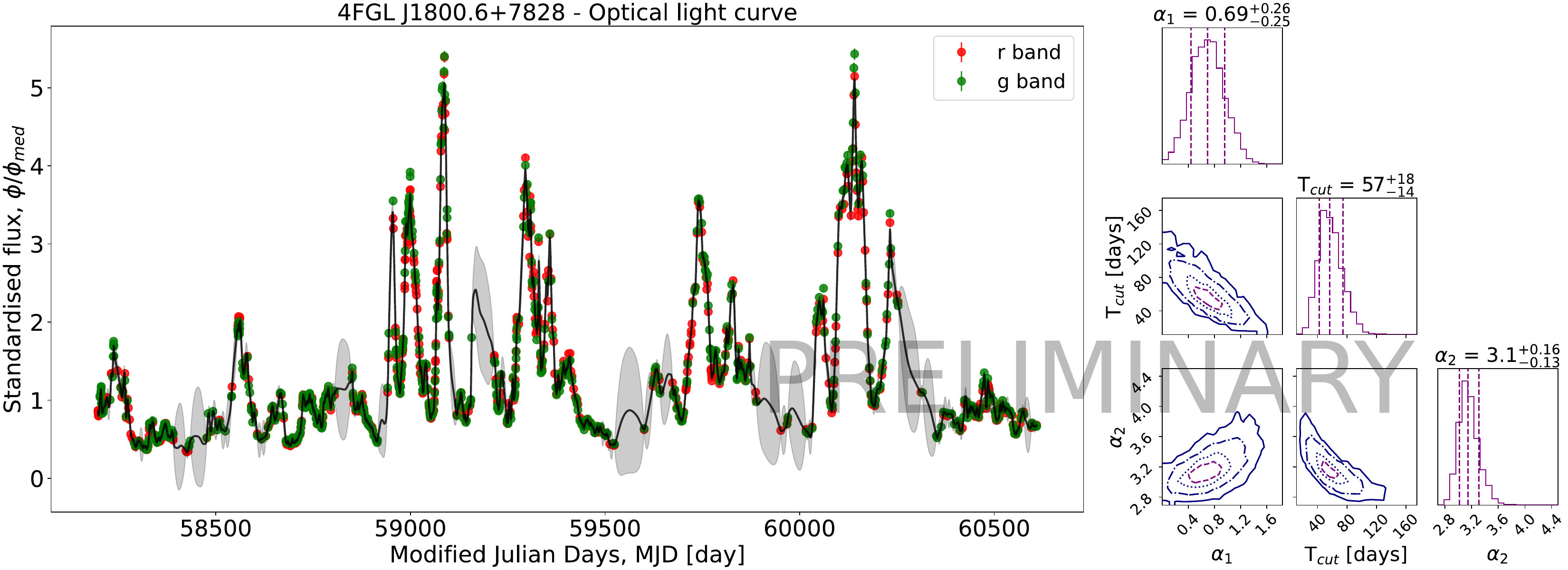}}
\caption[]{An illustration of the Gaussian process regression applied to ZTF observations of the blazar 4FGL\,J18000.6+7828, a low-frequency peaked BL Lac at $z=0.684$ also known as S5\,1803+78. \textit{Left:} Standardised time series constructed from g- and r-band observations. The black line and grey band show the best-fit and the associated 68\% credible interval obtained for a broken-power-law power spectral density.\cite{2025MNRAS.539.1775L} \textit{Right:} Reconstructed best-fit values of the power-spectral-density parameters: the low- and high-frequency indices ($\alpha_1$ and $\alpha_2$) and the time scale in days corresponding to the frequency break.}
\label{fig:GP}
\end{figure}

Figure~\ref{fig:GP} shows an example of such a standardised optical time series. We infer the properties in the Fourier space underlying this irregularly sampled and observationally uncertain time series using an implementation of Gaussian process regression developed for non-jetted active galactic nuclei observed in X-rays.\cite{2025MNRAS.539.1775L} Surprisingly, the power spectral density shows a break in the Fourier space, at a timescale consistent with that expected for a billion-solar-mass black hole accreting at 10\% of the Eddington rate.\cite{2019Galax...7...28R} This reconstruction is confirmed by the gamma-ray light curve, albeit with lower resolution due to the greater measurement uncertainties in the \textit{Fermi}-LAT fluxes. If confirmed in a larger sample, this promising result could provide an important observational link between accretion and ejection around supermassive black holes.

\begin{figure}
\centerline{\includegraphics[width=0.97\linewidth]{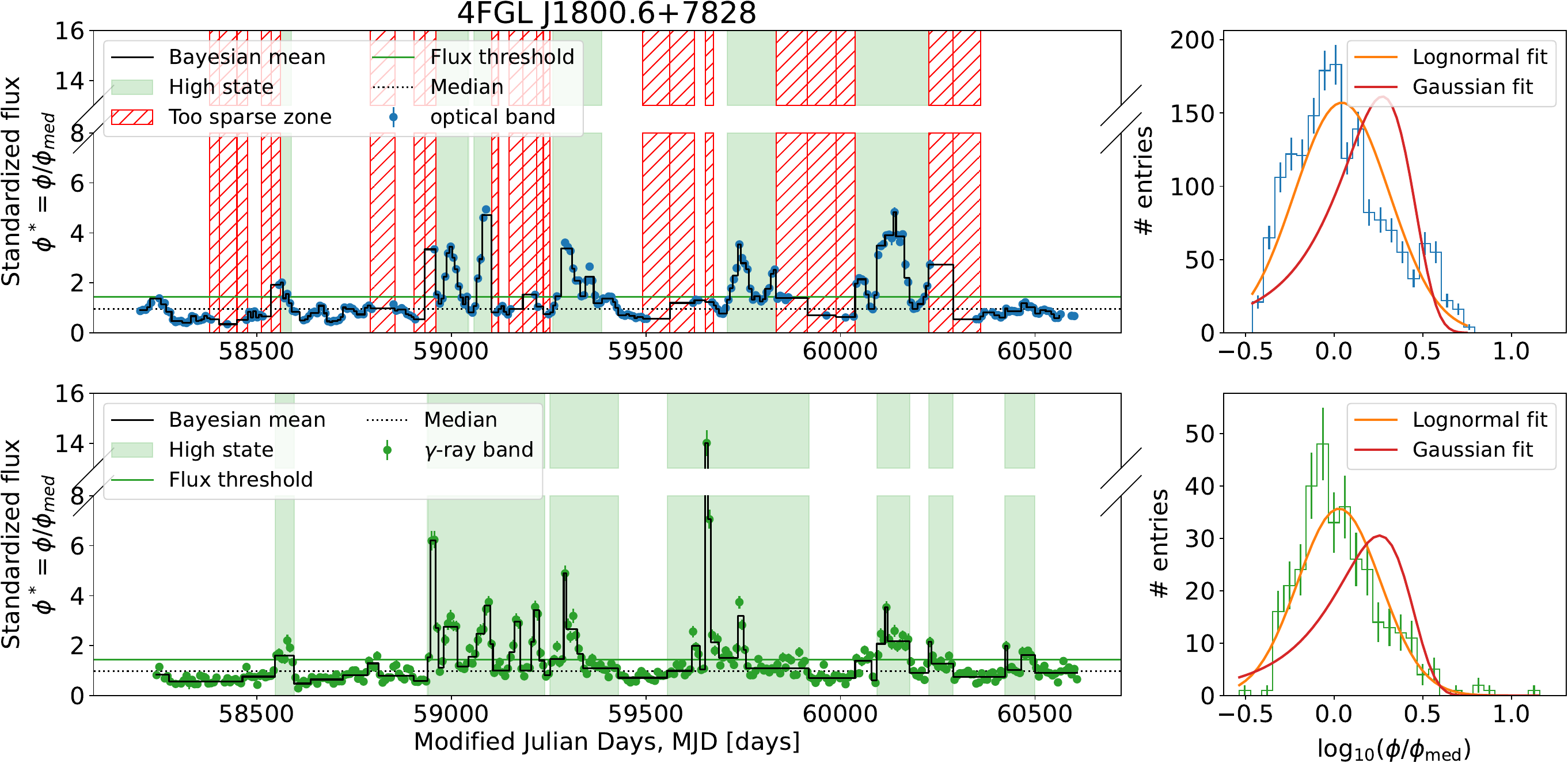}}
\caption[]{An illustration of the offline outburst-period identification algorithm applied the optical (\textit{top}) and gamma-ray (\textit{bottom}) observations of a blazar over a seven-year timespan. \textit{Left:} Periods hashed in red contain fewer than two points per month and are not used in the analysis. The black lines show the mean standardised flux obtained during consecutive Bayesian blocks. Series of blocks with rising and then decreasing flux are merged into super-blocks. Super-blocks containing a block flux above the threshold, shown as a horizontal green line set at the 75th percentile of the flux distribution, are identified as outburst periods and shaded in green. \textit{Right:} Distribution of the logarithm of the flux in each band. \textit{Adapted from Refs.\cite{HamoInPrep,MikhnoThisProc}.} }
\label{fig:BB}
\end{figure}

Standardised time series in the optical and gamma-ray bands are shown for the blazar used as an example in Fig.~\ref{fig:BB}. An experienced reader may note the similarity in the temporal structures observed in these bands over timescales of a few weeks to months, despite the fact that the bands differ by more than eight orders of magnitude in energy. Using a zero-normalised cross-correlation metric,\cite{2022arXiv221200476C} we measure a correlation of more than $6\sigma$ with a time lag compatible with zero between the optical and gamma-ray bands. This is based on the assumption that each time series is described as a memoryless white noise process under the null hypothesis (case in which Ref.~\cite{HamoInPrep} was able to establish an analytical expression for the significance). Considering instead as null hypothesis a pink-noise process, in line with the low- to mid-frequency behaviour in Fig.~\ref{fig:GP}, the significance of the correlation is conservatively estimated to more than $3\sigma$. This tentative correlation is observed to be linear rather than quadratic for this blazar. Such a behaviour is predicted by one-zone leptonic radiative models, enabling us to infer gamma-ray emission due to inverse Compton scattering of electrons off an external photon field rather than off the synchrotron field emitted by the same electrons in optical bands. The physics behind these time series is not only rich in detail but also in overview. Unbinned fitting of the flux distributions reveals a significant preference for a log-normal probability density function over a Gaussian distribution. This suggests that a multiplicative process, again possibly connected to accretion,\cite{2019Galax...7...28R}  is favoured as a variability-driving mechanism over the mere addition of `regular' random variables modelling independent emission regions. As indicated in Ref.~\cite{MikhnoThisProc}, we are extending these tests to heavy-tailed models suggested in additive scenarios involving Pareto-distributed bursts in jets and magnetic reconnection events.\cite{2012A&A...548A.123B}

As mentioned previously, our goal is to not only constrain the properties of the selected blazars and their emission mechanisms, but also to evaluate the potential for triggering very-high-energy gamma-ray observations with the CTAO based on large-scale optical surveys. To identify high-emission states, we use in Fig.~\ref{fig:BB} a decomposition into Bayesian blocks,\cite{2013ApJ...764..167S} which are then grouped into super-blocks via a hill-climbing method.\cite{2019ApJ...877...39M} These outburst periods are tagged for reference, as shown by the green zones in Fig.~\ref{fig:BB}. In Ref.~\cite{HamoInPrep}, we demonstrate how an implementation in Fink that accounts for the flux level and persistence of an  elevated optical state enables the generation of online alerts. Setting the trigger threshold at the 90th percentile of the optical flux in the online analysis rather than the reference 75th percentile used in the offline analysis yields a gamma-ray outburst purity of ${\sim}\,70\%$, at the cost of about 10\% efficiency, as evaluated on a sample of 86 sources dominated by low-frequency peaked blazars. On average, alerts are issued 10 days before the peak gamma-ray flux. By grouping alerts corresponding to the same optical outburst state, fewer than one alert per source per year is sent by Fink, resulting in an acceptable rate when tracking several hundred high-redshift blazars ($z>1$) with the CTAO transients handler.

\section{Discussion and conclusion}

In these proceedings, we explored the synergies between the Vera C. Rubin Observatory and the Cherenkov Telescope Array Observatory. Despite being separated by more than eight orders of magnitude in wavelength, these two observatories are both transient-detection machines. The former is about to issue optical alerts at a rate of several million per night, while the latter will be able to track the few gamma-ray outbursts per night or per week that represent the tip of the iceberg at the energy frontier.

Due to the high rate of optical alerts issued, the intermediaries required for this monitoring are the seven brokers selected by Rubin. We examined the potential of one of these brokers, Fink, for monitoring blazar outbursts. We demonstrated how developing scientific data processing modules within Fink, alongside filters that select the most promising events, enables monitoring of a sample of pre-selected gamma-ray sources. Alerts issued on Fink's dedicated channel, via the Apache Kafka protocol also used for GCN notices, will prove particularly useful for discovering distant blazar outbursts at $z>1$. These alerts are already available in the Northern Hemisphere for blazars detected by ZTF, together with gamma-ray alerts from FLaapLUC.\cite{Lenain:2017okf} Our future work will focus on identifying blazar candidates in optical survey data that do not have direct counterparts.

Blazars are not the only sources of interest for such a synergy. Jetted tidal disruption events, gamma-ray bursts, supernova shock breakouts and the electromagnetic counterparts of gravitational waves also warrant dedicated investigation. While initial implementations for certain classes of these sources are already available in Fink, we encourage the wider astroparticle physics community to reach out and participate in the hackathons organised to develop filters that best suit their needs. The Fink broker does not evolve in isolation. A process of sharing tools developed by the Fink, Lasair and AMPEL brokers is already underway within the framework of the European ACME programme, enabling developments on one platform to benefit the others.

This work on the connections between optical and gamma-ray observations should not overshadow the importance of the rest of the multi-wavelength and multi-messenger landscape. The issuance or monitoring of alerts by actors covering the entire spectrum is desirable for many non-thermal sources. This is particularly true for wide-field X-ray and gamma-ray observatories, as well as future radio observatories. The maturation of time-domain astronomy, spearheaded by the Legacy Survey of Space and Time, provides an unprecedented opportunity to share our discoveries in real time and coordinate our observation campaigns.

\section*{Acknowledgments}

\small{The authors acknowledge feedback on Rubin from J.~Bregeon, on the manuscript from E.~Prandini and the CTAO internal reviewer, M.~Meyer, and on Gaussian process regression from M.~Lefkir and P.~Uttley. The text of this paper is entirely original and was drafted in full by the authors. DeepL was used to refine the readability of the prose. The authors verified that no meaning was altered during this process. This work was conducted in the context of the CTA Consortium. We gratefully acknowledge financial support from the agencies and organisations listed here: \url{https://www.ctao.org/for-scientists/library/acknowledgments/}. This work was made possible by Institut Pascal at Universit\'e Paris-Saclay with the support of the program ``Investissements d’avenir" ANR-11-IDEX-0003-01, the P2I axis of the Graduate School of Physics of Universit\'e Paris-Saclay, as well as IJCLab, CEA, IAS, OSUPS, and APPEC.}

\section*{References}
\bibliography{moriond}

\end{document}